\documentclass[
    ,final            % use final for the camera ready runs
      ]
        {aipproc}
	
\layoutstyle{8x11double}

\begin{document}

\title 
      [Gamma rays from galaxy clusters]
      {Gamma ray emission and stochastic particle acceleration in galaxy
      clusters}


\author{G. Brunetti}{
  address={INAF Istituto di Radioastronomia, via P.Gobetti 101, 40129,
  Bologna, Italy},
  email={brunetti@ira.inaf.it},
}
\iftrue
\author{P. Blasi}{
address={INAF Osservatorio Astrofisico di Arcetri, Firenze, Italy and
Theoretical Astrophysics, Fermi National Accelerator Laboratory,
Batavia, USA},
}
\iftrue
\author{R. Cassano}{
address={INAF Istituto di Radioastronomia, via P.Gobetti 101, 40129,
    Bologna, Italy},
}
\iftrue
\author{S. Gabici}{
address={Dublin Institute for Advanced Studies, 31 Fitzwilliam Place,
Dublin, Ireland}
}

\fi
%\iftrue
%\author{Arno Mittelbach}{
%  address={Zedernweg 62, 55128 Mainz, Germany},
%  email={arno@mittelbach-online.de},
%}
%
%\author{D. P. Carlisle}{
%  address={Willow House, Souldern},
%  email={david@dcarlisle.demon.co.uk},
%  homepage={http://www.dcarlisle.demon.co.uk},
%  altaddress={When I go to work: NAG Ltd, Oxford}
%}
%\fi

% \copyrightholder{Acoustical Scociety of America}
\copyrightyear  {2001}

\begin{abstract}
FERMI (formely GLAST) will shortly provide crucial
information on relativistic particles in galaxy clusters. 
We discuss non-thermal emission 
in the context of general calculations in which
relativistic particles (protons and secondary electrons due to
proton-proton collisions) interact with MHD turbulence generated in
the cluster volume during cluster mergers.
Diffuse cluster-scale radio emission (Radio Halos) and hard X-rays
are produced 
during massive mergers while gamma ray emission, at some level,
is expected to be common in galaxy clusters.
\end{abstract}

\date{\today}

\maketitle

\section{Introduction}

Clusters of galaxies contain $\approx 10^{15}$ M$_{\odot}$ 
of hot ($10^8$ K) gas, galaxies, dark matter and non-thermal components.  

The origin of non-thermal components 
is likely connected with the cluster formation process : 
a fraction of the energy dissipated during cluster mergers is expected to be 
channelled into the acceleration of particles via shocks 
and turbulence that lead to a complex 
population of primary electrons and protons in the IGM 
(e.g., \cite{enss98, sara99, blasi01, gb01, gb04, petro01, min01, ryu03, 
gb+laz07, pf08}).
Theoretically relativistic protons are expected to be the dominant
non-thermal particle components since they have long life-times
and remain confined within galaxy clusters 
for an Hubble time (e.g. \cite{blasi07} and ref. therein). 
Confinement enhances the probability to have proton-proton (p-p) collisions 
that in turns give gamma ray emission via decay of the neutral pions produced 
during these collisions \cite{voelk96, bere97}.
p-p collisions also inject secondary electrons that 
give synchrotron and inverse Compton (IC) emission whose relevance 
depends on the proton content in the IGM \cite{blasi99}.
The inter-galactic medium (IGM) is expected to be turbulent 
at some level and MHD turbulence 
can re-accelerate both primary and secondary particles 
via second order Fermi mechanisms. 
Turbulence is naturally generated in cluster mergers 
(e.g., \cite{ricker+sara01, dolag05, vazza06, iapichino08}) and the resulting
particle re-acceleration process should enhance the synchrotron 
radio emission and the IC hard X--ray emission by orders of magnitude.

The energy content in the form of relativistic protons in the IGM 
is still poorly constrained since present gamma ray observations 
can provide only upper limits to the gamma ray 
emission from galaxy clusters (\cite{reimer03}; also Reimer and Perkins, this
conference).
On the other hand radio observations of Mpc-sized diffuse synchrotron 
emission from galaxy clusters provide crucial information on the 
relativistic electron component (e.g. \cite{feretti03, ferrari08}).

The FERMI gamma-ray space telescope will
shortly allow to measure (constrain) the energy content of relativistic 
protons in the IGM. 
For this reason, starting from present understanding of non thermal
components in galaxy clusters, we discuss expectations for 
gamma ray emission.

\section{Cluster-scale radio emission}

In this Section we outline our starting point to model 
non thermal (including gamma ray) cluster emission.

The most prominent examples of diffuse cluster 
emission are giant Radio Halos: Mpc-scale diffuse synchrotron sources
at the centre of a fraction of massive and merging galaxy 
clusters \cite{feretti03, ferrari08}.
This Mpc-scale radiation may originate from {\it secondary electrons}
injected by collisions between relativistic and thermal protons in
the IGM (e.g. \cite{dennison80, blasi99}),
alternatively 
extended radio emission may originate from relativistic electrons
{\it re-accelerated} {\it in situ} by various mechanisms associated with
the turbulence in massive merger events (e.g. \cite{gb01, petro01, 
fujita03}).
These two processes likely happen at the same time and
a unified scenario that models both the injection and re-acceleration
of {\it secondary electrons} and {\it primary particles} due to
MHD turbulence has been investigated by \cite{gb+blasi05}.

Radio Halos are not common: although a fairly large number of 
clusters has adequate radio follow up, they are presently detected 
only in a fraction of massive and merging clusters \cite{gg99,
buote01, cassano08, venturi08}.
Studies based on the analysis of present X--ray selected cluster samples 
with radio follow up allow to conclude that the fraction of clusters 
with Radio Halos depends on cluster X-ray luminosity\footnote{the 
fraction is$\approx 30\%$ 
for $L_X > 8\cdot 10^{44}$ erg s$^{-1}$ clusters and 
$\leq$ 10\% for clusters with 
$L_X \approx 3\cdot 10^{44}-8\cdot 10^{44}$ erg s$^{-1}$;
Montecarlo approaches show that the two fractions differ 
at $\approx 4 \sigma$ level \cite{cassano08}}, and that
clusters have a bimodal behaviour in the 1.4 GHz radio luminosity ($P_{1.4}$)
-- soft X--ray luminosity ($L_x$) plane
with Radio Halo clusters being one order of magnitude more radio
luminous than upper limits for 
clusters with no Radio Halos (\cite{gb07}; Fig.~1).

These facts suggest that Radio Halos are 
{\it transient} phenomena\footnote{this means that assuming that all massive
clusters may host Radio Halos during their life, the life-time of
Radio Halos must be much shorther, $< 1$ Gyr, than the cluster 
life-time.} connected with cluster mergers
and that some threshold in the mechanism
for the generation of these sources should come into play.
Unless we admit the possibility of strong  
dissipation of the magnetic field in clusters, these properties cannot
be easily understood in the case that the continuous injection of
{\it secondary} electrons in the IGM plays the major role in the origin
of these sources.
In this case - indeed - Radio Halos should be {\it long living phenomena}
and common, and some general $P_{1.4}$--$L_x$ trend would be predicted 
for all clusters (e.g. \cite{min01, dolag06, pf08}).
In particular, the bimodality in Fig.~1 and the substantial lack of 
clusters between the Radio Halo and radio quiet regions implies that 
the bulk of the magnetic energy in clusters should be 
dissipated in a time--scale of only $\approx 0.1-0.2$ Gyr which is 
challenging to reconcile with present understanding
of cluster magnetic fields (e.g. \cite{subramanian06}\footnote{
even in the (worst) case of simply decaying MHD turbulence, the energy 
density of the rms field decreases with time only (about) linearly
and $\geq$ 5 eddy turnover times ($\approx$ Gyr) are required to allow 
dissipation of the bulk of the magnetic field 
(Fig.2 in \cite{subramanian06})}).

On the other hand, the emerging observational picture supports 
the idea that turbulent re-acceleration of relativistic electrons 
may play a role in the formation 
of Radio Halos in connection with cluster mergers. 
In particular it is interesting to note that in the context of this
scenario radio-quiet clusters are expected to evolve into
Radio Halo clusters (and vice versa) in a time--scale of the order
of the electron acceleration time-scale, $\approx$0.1-0.2 Gyr.
Spectral studies are also important : the discovery of Radio Halos
with steep spectrum \citep{gb08} and the synchrotron cut off
that is found 
in the spectrum of the Coma Radio Halo \citep{schl87, thier03}
imply a maximum energy in the spectrum of the emitting electrons 
at energies $\approx$GeV suggesting that the mechanism responsible for the
acceleration of electrons in the IGM is poorly efficient, consistent 
with turbulent acceleration.

\begin{figure}
\caption{
Distribution of galaxy clusters in the
$P_{1.4}$--$L_x$ plane. Filled dots are galaxy clusters at
z$\geq$0.2 (from GMRT-sample and from the literature), empty dots
are clusters at lower redshift reported to highlight the
$P_{1.4}$--$L_x$ correlation (solid line).
Upper limits (represented by the arrows) are GMRT clusters with no
hint of cluster-scale emission and their distribution should be
compared with that of clusters at similar redshift (filled dots).}
  \includegraphics[height=.3\textheight]{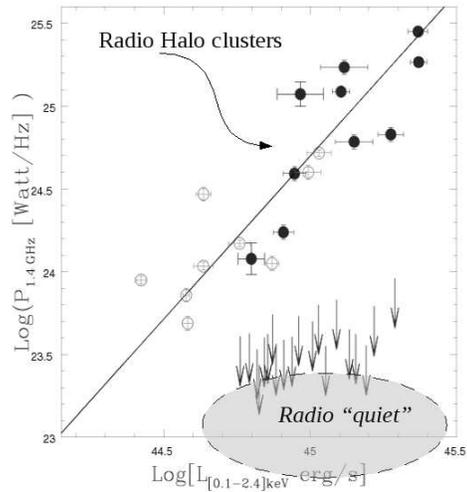}
\end{figure}

\begin{figure}
\caption{
Broad band spectrum produced within $R < 1$ Mpc from a Coma-like cluster.
{\bf Upper panels}: Synchrotron (left, SZ decrement at high
frequencies is not taken into account), and IC and $\pi$-o emission
(right) calculated at t=0.5 Gyr from the injection of MHD
turbulence in the IGM (the energy injected in Alfven modes between
t=0-0.5 Gyr is $\approx$3\% of the thermal energy). 
{\bf Lower panels}: Synchrotron (left) and IC
and $\pi$-o emission (right) calculated at t=1 Gyr after dissipation
of turbulence in the IGM.
In all panels calculations
are shown assuming a ratio between the energy density of
relativistic and thermal protons = 1\% (dashed lines), 0.5\% (dotted lines) 
and 0.3\% (solid lines) at t=0 (with proton spectrum $\delta
=2.2$) and a central cluster-magnetic field $B_o=2 \mu$G.
For the sake of completeness we show radio data, BeppoSAX data 
and EGRET upper limit for the Coma cluster (\cite{gb+blasi05} and
ref. therein) the recent VERITAS upper limit (Perkins, this meeting) and
the approximate sensitivity after 1 yr of FERMI (dashed). 
}
\includegraphics[height=.42\textheight]{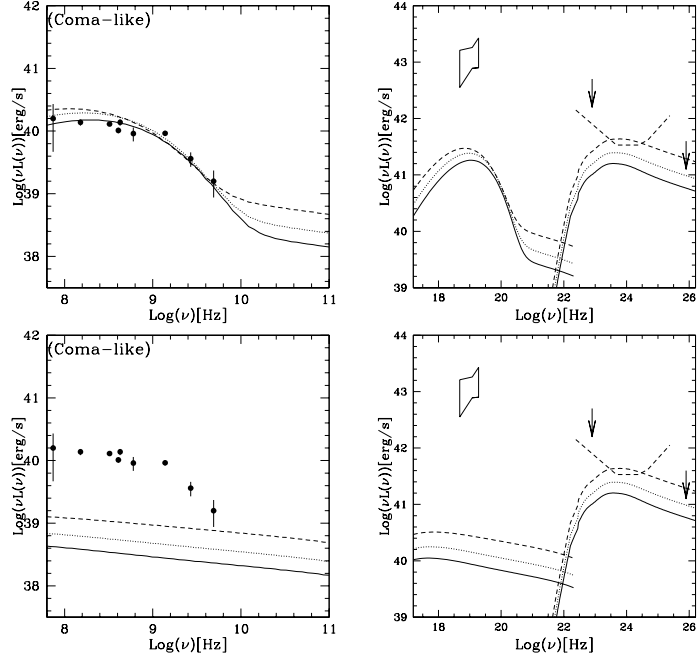}
\end{figure}

\section{Emission from galaxy clusters}

In this Section we calculate non-thermal (multifrequency) emission
from galaxy clusters under the assumptions that MHD turbulence
plays a role in the
particle acceleration process during cluster mergers.
We do not include the contribution to the non thermal
emission from fast electrons accelerated at shock waves that develop 
during cluster mergers and accretion of matter (see Pfrommer, this
conference).

As already mentioned, it is believed that clusters are {\it reservoir}
of relativistic protons that accumulate in the IGM during cluster
life-time (e.g. \cite{blasi07}).
Thus we model the re-acceleration of
relativistic particles by MHD turbulence in the most simple
situation in which only relativistic protons
are initially present in a turbulent IGM. These protons generate
secondary electrons via p-p collisions and in turns secondaries (as
well as protons) are re-accelerated by MHD turbulence. 
Following \cite{gb+blasi05} we restrict to the case of
Alfven waves\footnote{an additional possibility is given
by magnetosonic waves \cite{cassano+gb05, gb+laz07}}
and calculate the spectrum of particles 
and MHD waves and their evolution with time by solving a set
of coupled equations
that give the spectrum of electrons, $N_e^-$, positrons, $N_e^+$,
protons, $N_p$, and waves, $W_k$ :

\begin{eqnarray}
{{\partial N_e^{\pm}(p,t)}\over{\partial t}}=
{{\partial }\over{\partial p}}
\Big[
N_e^{\pm}(p,t)\Big(
\left|{{dp}\over{dt}}_{\rm r}\right| -
{1\over{p^2}}{{\partial }\over{\partial p}}(p^2 D_{\rm pp}^{\pm})
\nonumber\\
+ \left|{{dp}\over{dt}}_{\rm i}
\right| \Big)\Big]
+ {{\partial^2 }\over{\partial p^2}}
\left[
D_{\rm pp}^{\pm} N_e^{\pm}(p,t) \right] \nonumber\\
+ Q_e^{\pm}[p,t;N_p(p,t)] \, ,
\label{elettroni}
\end{eqnarray}

\begin{eqnarray}
{{\partial N_p(p,t)}\over{\partial t}}=
{{\partial }\over{\partial p}}
\Big[
N_p(p,t)\Big( \left|{{dp}\over{dt}}_{\rm i}\right|
-{1\over{p^2}}{{\partial }\over{\partial p}}(p^2 D_{\rm pp})
\Big)\Big]
\nonumber\\
+ {{\partial^2 }\over{\partial p^2}}
\left[ D_{\rm pp} N_p(p,t) \right] \, ,
\label{protoni}
\end{eqnarray}

and 

\begin{eqnarray}
{{\partial W_{\rm k}(t)}\over
{\partial t}} =
{{\partial}\over{\partial k}}
\left( k^2 D_{\rm kk} {{\partial}\over{\partial k}} \left[
{{W_k(t)}\over{k^2}} \right] \right)
- \Gamma(k) W_{\rm k}(t) 
\nonumber\\
+ I_{\rm k}(t), 
\label{turbulence}
\end{eqnarray}

where $|dp/dt|$ marks radiative (r) and Coulomb (i) losses, 
$D_{pp}$ is the particle diffusion coefficient in the momentum
space (and depends on the wave spectrum 
$W_{\rm k}$), $Q_{e}^{\pm}$ is
the injection term of secondary leptons due to p-p collisions 
(and depends on $N_p$), $D_{kk}$ is the diffusion coefficient in 
the wavenumber space,
$I_k$ is the injection rate-spectrum of Alfven waves at resonant
scales, and $\Gamma$ is the damping rate of waves due to non-linear
resonance with thermal and relativistic particles ($N_p$ and $N_e$);  
details can be found in \cite{gb+blasi05}.

In the following we consider a simple model of galaxy cluster assuming 
that the energy densities
of relativistic protons at the beginning of reacceleration, $\epsilon_{p}$, 
and of the magnetic field, $B$, and the injection rate of Alfven 
waves during mergers, $I_{k}$, scale with thermal energy 
density, $\epsilon_{th}$ ($\epsilon_{p} \propto \epsilon_{th}$, 
$B \propto \epsilon_{th}$ and $\int I_{k} dk \propto \epsilon_{th}$).
An example of the expected broad band emission (synchrotron, IC, $\pi^o$
decay) is reported in Fig.~2 by adopting the spatial distribution
and physical parameters of the thermal IGM of the Coma cluster.
In the context of this model 
the non-thermal emission is a mixture of two main spectral
components: a long-living one that is emitted by secondary particles (and by
$\pi^o$ decay) continuously generated during p-p collisions, 
and a transient component that is due to the re-acceleration of
relativistic particles by MHD turbulence generated (and then
dissipated) in cluster mergers.
In order to highlight these components we calculate the non-thermal 
emission during a cluster merger, i.e. assuming a turbulent IGM,
(Fig.~2, upper panels) and 1 Gyr after turbulence 
is dissipated (Fig.~2, lower panels).
Lower panels show the long-living component of the non-thermal 
cluster emission, because relativistic protons, that in turns generate 
secondaries, lose energy on a long time-scale.
This long-living emission 
does not strongly depend on the dynamics of clusters but
only on the energy content
(and spectrum) of relativistic protons in the IGM (and on the magnetic
field in the case of the synchrotron radio emission).
On the other hand, the comparison between upper and lower
panels of Fig.~2 highlights the transient emission that is generated 
by short-living electrons reaccelerated by turbulence during cluster mergers.

We note that the results in Fig.~2 have the potential 
to reproduce the radio bimodality observed in galaxy clusters
(Fig.~1) : Radio Halos develop in connection with
particle re-acceleration due to MHD turbulence in cluster mergers
where the cluster-synchrotron emission is considerably
boosted up (upper panel),
while a fainter long-living radio emission from secondary electrons
is expected to be common in clusters (lower panel); the level of this
latter component must be consistent with the radio upper limits from 
radio observations of clusters with no Radio Halos.
IC hard X-rays are also produced in connection with Radio Halos, 
although the IC signal from re-accelerated {\it secondary} electrons 
is not expected to be very luminous 
(see discussion in \citep{gb+blasi05} for a comparison with the
case of re-acceleration of {\it primary relic} electrons).

An important point is that gamma ray emission 
is expected (at some level, depending on the content and
spatial distribution of relativistic protons) to be common
in galaxy clusters and not directly 
correlated with the presence of giant Radio Halos.
Cerenkov arrays already constrain the level of gamma rays from
Coma and few other nearby clusters (Fig.~2, Perkins et al., this meeting).
After $\approx$1 yr of operations
FERMI will reach adequate sensitivity in the energy 
range 0.1-100 GeV to start obtaining crucial constraints on
nearby clusters and hopefully to measure the energy content of relativistic
protons in these clusters.

\begin{theacknowledgments}
We acknowledge partial support from ASI-INAF I/088/06/0 and 
PRIN-INAF2007, and useful discussions with M. Giroletti.
\end{theacknowledgments}

% choose bibtex style depending on layout style and options used in
% sample:
\bibliographystyle{aipprocl} % if natbib is missing

{}

\end{document}